# Quality Assurance in the Context of Contemporary Software Practice

Stephen G. MacDonell, May 2022

*Introduction*

After the formal introduction of the Agile Manifesto in 2001 (Beck et al., 2001), previously conventional life-cycle models were at first slowly and then very rapidly augmented, or replaced outright, by approaches that emphasised rapid iterations or increments of idea-development-deployment cycles (Boehm et al., 2019a; Bruel et al., 2019). In the last decade this has been followed by adoption of 'continuous *' (or 'continuous everything') thinking (Fitzgerald and Stol, 2017), and rapid continuous data-driven software engineering (Gerostathopoulos et al., 2019; Konersmann et al., 2020). This was seen first in activities such as continuous integration and continuous delivery (CI/CD) of software features, releases, products and services. Subsequently, the traditional *development* phases of requirements, design, implementation and the like, and the often separately treated catch-all phase called *maintenance*, have been melded into a continuous deployment pipeline of new features alongside bug fixes, enhancements and refactorings (Cheriyan et al., 2020; Kersten, 2019).

As a result, the once-central notion of a software 'life cycle' now assumes less importance than an adaptive software process. Similarly, the Rethinking Project Management movement (Sauer and Reich, 2009; Svejvig and Andersen, 2015) and the growth of software-as-a-service (SaaS) have seen the structural frames of software 'projects' and even 'products' also lose traction, giving way to an approach in which software features and changes are delivered dynamically as needed as decoupled services. Formerly co-located teams are increasingly distributed/remote, and their work is increasingly global (Drechsler and Breth, 2019; Shameem et al., 2018). Agile and hybrid methods, and more importantly an agile mindset, are pervading other organisational functions (Digital.ai, 2021; KPMG, 2019), as creative or disruptive input models such as design thinking are adopted for product, and service, development (Corral and Fronza, 2018).

This is the reality of much contemporary software practice. While this dynamic reality could be perceived as presenting a threat to quality, it instead presents an opportunity to reconsider quality within a new operational context (Cheriyan et al., 2020).

*Quality Assurance is inevitably incomplete, out of date and/or imperfect*

It is important to acknowledge that, even before the advent of the shifts described above, for all but the simplest single-user, single-function software programs, quality assurance (QA) quickly became intractable. Even as early as 1986 the computing community acknowledged that "…software entities are more complex for their size than perhaps any other human construct…" and it was impossible for any given individual to understand everything there was to know about a software system (Brooks, 1986). There are simply too many unknowns to be able to anticipate in advance, especially given the potential for changes in the environment that may affect the operation of software in a system, as well as in terms of unexpected user behaviours, both malicious and unintended. QA has therefore long been recognised as a containment exercise based on informed risk management and mitigation, not as a cast-iron and infinite guarantee of functional and non-functional outcomes.

The consequences that might arise from constrained QA therefore do inevitably occur, ranging from the inconvenient or mildly annoying to the mission- or life-threatening. Even in high-risk, high-cost endeavours mistakes are made, and the results can be catastrophic: some of the more infamous examples are well-known across healthcare (Therac-25)[1], space exploration (Ariane-5[2]), and avionics (Boeing 737-Max)[3]. The instances are so numerous, in fact, that they have sustained a frequently published *Risks Digest* for the last 37 years. Since 1985 the *Forum on Risks to the Public in Computers and Related Systems* moderated by Peter G. Neumann has reported a digest of problems that have occurred in system development, deployment and use, detailed across 90+ issues of the digest each year. Tens of thousands of risks, errors, faults and their consequences have been identified in that time (Risks Digest, n.d.).

This is all not to say that we should simply give up on QA, but rather, that we should accept that the complexity of contemporary software systems is such that we need to be pragmatic about what QA can and cannot do. In particular, placing reliance on a single QA strategy, method or technique is highly risky.

Given the above it is quite appropriate, then, that even industry standards include disclaimers regarding their use, and offer no guarantees regarding system behaviours or the outcomes of system use. IEEE Standard 1012-2017 (IEEE, 2017, p.17) notes: "The dynamics of complex systems and the multitude of different logic paths available within the system in response to varying stimuli and conditions demand that the V&V [verification and validation] effort examines the correctness of the system for each possible variation in conditions. The ability to model complex, real-world conditions will be limited, and thus, the V&V effort examines whether the limits of the modeling are realistic and reasonable for the desired solution. The unlimited combination of system conditions presents the V&V effort with the challenge of using a finite set of analytical, test, simulation, and demonstration techniques to establish a reasonable body of evidence that the system is correct."

An additional complication regarding standards is that there are multiple versions and variants, which may or may not be applied to any given software/system development and deployment effort. Prominent in this space are the standards produced by international bodies the IEEE[4], the International Electrotechnical Commission (IEC)[5] and ISO[6], the International Organization for Standardization. Specific US government agencies, including the Departments of Energy[7] and Defense[8], have set their own standards for systems that are to be deployed in their particular operational contexts. There are yet further jurisdictional entities such as the UK's national standards body BSI[9], Europe's CEN and CENELEC[10], and Standards Australia[11], who have locally applicable standards. And in some cases, standards are co-badged, where development has been a collaborative effort across standards bodies, or where a standard developed by one authority is then endorsed or accepted by another. (See also Appendix 1.)

---

[1] https://catless.ncl.ac.uk/Risks/9/45#subj7.1
[2] https://catless.ncl.ac.uk/Risks/18/28#subj9.1
[3] https://catless.ncl.ac.uk/Risks/31/17#subj1.1
[4] https://www.ieee.org//
[5] https://www.iec.ch/homepage
[6] https://www.iso.org/home.html
[7] https://www.energy.gov/
[8] https://www.defense.gov/
[9] https://www.bsigroup.com/
[10] https://www.cencenelec.eu/
[11] https://www.standards.org.au/

In addition, and as noted above, technologies and methodologies change, sometimes continuously but often in sequences of relative stability followed by a 'spike' or a major, rapid shift. Operational contexts adjust to meet strategic opportunities, competitive pressures, or legislative moves. Commerce and science change as new knowledge informs new practice. It is inevitable, then, that approaches to quality assurance need to change with them (Yoder et al., 2015).

This may not be straightforward, however. Carrozza et al. (2018) note that, in the context of today's large-scale mission-critical software-intensive systems (pp.100, 113), "[i]nnovations in software quality practices are hard to introduce… because of skepticism of managers, who are often anchored to consolidated industrial processes where software is seen just as an intangible "add-on" to the concrete system [a cultural barrier also noted by Islam and Storer (2020) in their study of an avionics company engaged in safety-critical system engineering, and by Smith et al. (2019) at NASA (p.6): "The word Agile has a dirty feel to it around the halls of NASA IV&V."]. In such contexts, the generic perception of the importance of software quality needs to be supported in quantitative ways… Innovation in software quality management is not simply a matter of applying established techniques and adopting proper tools."

Furthermore, given the range and speed of change of some of those influencing factors described above, incremental approaches to development, deployment *and* quality assurance are increasingly worthy of consideration. The recent release of IEEE Standard 2675™-2021 (IEEE, 2021) focused on the building, packaging and deployment of reliable and secure systems under a DevOps way of working is illustrative of this need for ongoing reflection, revision and evolution of approaches.

Finally, it is important to note that the human-in-the-loop remains a key component of many software systems, across all activities including those influencing its design, build, test, V&V and operation. At all of those touchpoints there is scope for manual scrutiny of what goes into and what comes out of each, whether that be related to the quality of the underpinning science, the specification of the algorithms, the review of code, the verification of outputs or the interpretation of the results – and thereby fully utilising a "…finite set of analytical, test, simulation, and demonstration techniques…" as noted above. This creates a 'dynamic and collective QA' approach whereby the limitations of one strategy, method or technique can be offset by the capabilities of another; where adaptive development and deployment processes intentionally build in contingencies and redundancies to cope with the inevitable misses; and where early warning systems trigger lesser consequences but help to minimise the likelihood of catastrophic failures.

*Contemporary approaches to building software and software-intensive systems*

Software process adaptation has been commonplace for many years. Even the so-called waterfall model of development, with its apparently rigid linear sequence of phases with hard sign-offs at the end of each, was not originally designed as such – it was always anticipated that projects would use cycles of iteration and information flows back and forth as needed (Royce, 1987). IEEE Standard 1012-2017 (IEEE 2017, p.15) also notes that the "…standard is compatible with all life cycle models (e.g., system, software, and hardware); however, not all life cycle models use all of the life cycle processes described in this standard… The user of this standard may invoke those life cycle processes and the associated V&V processes that apply to the project."

It continues (p.21): "If a project uses only selected life cycle processes, then conformance to this standard is achieved if the minimum V&V tasks are implemented for all of the associated life cycle processes selected for the project… Specific development methods and technologies (such as automated code generation from detailed design) may eliminate development steps or combine several development steps into one; therefore, a corresponding adaptation of the minimum V&V tasks is permitted and is documented in any claim of conformance to this standard."

Importantly, it includes the following under the heading "1.8 Disclaimer" (p.21): This standard establishes minimum criteria for V&V processes, activities, and tasks. However, implementing these criteria does not automatically ensure conformance to system or mission objectives, or prevent adverse consequences (e.g., loss of life, mission failure, loss of system safety or security, or financial or social loss). Conformance to this standard does not absolve any party from any social, moral, financial, or legal obligations."

The above paragraphs acknowledge three things: that 'life cycles', now processes and pipelines, vary; that the verification and validation activities that support those processes should therefore also vary in line with them; and that conformance to a standard does not by definition prevent failure or loss.

Industry surveys of (agile) processes and practices have been undertaken for more than a decade, the most widely known being the VersionOne Inc. then Digital.ai Annual State of Agile, in its 15$^{th}$ iteration in 2021 (Digital.ai, 2021). Others have a more specific focus, on agile transformation or on agile metrics, for instance (KPMG, 2019; Wolpers, 2021). These have been complemented in the last five years by an international survey led and conducted by academics in an effort to lend greater analytic rigour to such efforts. Known as the HELENA Project (Hybrid dEveLopmENt Approaches in software systems development), the study was undertaken over three phases, culminating in a single multi-language survey distributed across 20+ countries by 60+ international collaborators.

Detailed results of the HELENA study have been reported in a number of peer-reviewed publications (e.g., Klünder et al., 2019; Kuhrmann et al., 2019; Kuhrmann et al., In Press; Tell et al., 2019; Tell et al., 2021) and we summarise some of the key findings here:

- Process (and tool) variation, adaptation and customisation are commonplace.
- Decisions made around processes and practices emphasise fitness for purpose.
- There can be variation from one development effort to another, and certainly across teams, within the same organisation. Experimentation is encouraged.
- Teams and organisations that are distributed across nations and time zones are prevalent.
- There are some adopters of prescriptions – process recipes – but this is often early in the move to agile/hybrid, or to give the impression of using more contemporary methods.
- More common is the use of multiple hybrid process variants: Scrumbut, Scrumban, Water-Scrum-Fall, and many others.

The latter methods are predominant, even in traditionally process-heavy sectors, requiring new ways of thinking about V&V: "The stakes are quite different if you have a V&V (validation and verification) phase of a few weeks to prepare for the next release, as in the old world…, and in the brave new world of deploying this morning's change in the afternoon for the millions of users of your Web-based offering." (Bruel et al., 2019, p.v). Cheriyan et al. (2020) propose an 'Agile continuous integration and delivery—quality assurance (ACID-QA) model' that seeks to assure quality through: Continuous Delivery Readiness, QA in the Pipeline, Quantitative and Causal Analysis, Learning and Development, and Team Culture.

This type of thinking regarding V&V is pervading all sectors. In the automotive industry (Puntigam et al. 2021, p.471) "…[a] new approach is required to enable continuous vehicle-level verification and validation in all phases of the vehicle development process.", a point made previously by Tahera et al. (2019). Industry 4.0 manufacturing "…requires an agile and dynamic production process to be competitive in the market" (Ugarte Querejeta et al., 2020, p.205). Liang et al. (2021) note that (p.384) "…software V&V development of nuclear power plants will focus on… [a]gile testing… automation… [and s]ervicing to make software a service." In maritime traffic systems Reiher and Hahn (2021) promote continuous V&V using scenarios as (p.1) "…classical methods for safety verification are no longer suitable".

*Towards continuously agile (independent) verification and validation*

Can contemporary development and deployment practices and verification and validation be reconciled and so used together? Could continuously agile V&V even improve quality outcomes?

*MBSE, agile methods and V&V*

Model-Based Software/Systems Engineering (MBSE) offers one approach that can be used to formally verify and validate operational, functional and technical aspects of system behaviour.

Borky and Bradley (2019) leverage their experience in aerospace and defense to propose a Model-Based System Architecture Process (MBSAP) that employs virtual and physical prototypes to (p.405) "…allow an earlier start on V&V while the details of the system design are evolving, and subsequent data from [integration and test] I&T activities then serve to refine and confirm the modeled results." They encourage the use of incremental (and potentially spiral) 'development builds' to (p.408) "…add relatively small capability increments, verify problem fixes, and support ongoing engineering activities such as verifying the results of trade studies. A series of development builds generally leads up to a system build."

With respect to V&V they note the need for a range of assessments methods, including testing, demonstration, inspection and analysis, all supported by experimentation, certification, modelling and simulation, comparison and implication, with the human-in-the-loop playing a central role in exercising the prototype builds to identify issues and required fixes. Borky and Bradley (2019) note that V&V is most effective when it is 'seamless', meaning it minimises the seams between those involved (p.417-418), "…especially the acquiring organization and the contractor or contractors, who design, build, integrate, and test the system… [W]ith complex, high-technology systems… exhaustive testing of every system behavior under every operating condition is prohibitive in cost and schedule. Effective test planning is therefore based on identifying stressing test cases whose measured outcomes yield adequate confidence that the full requirements space has been evaluated."

Baduel et al. (2108) report on its use at Bombardier Transportation in producing a portfolio of railway products. While the authors promote the potential of MBSE they also note (Baduel et al., 2018, p.134, 137): "Then they [the models] are validated by the domain experts to ensure that the system model representation complies with the specification of the real-world system and the system requirements… The domain expert plays a crucial role in validating the system models' content based on his [sic] own experience of the real-world system represented by system models."

They further acknowledge that V&V can be performed in a variety of ways and that it will be increasingly important to look at opportunities for V&V automation to ensure efficiency as well as accuracy – rule-based approaches to validation are not tractable from a time and skills perspective given the complexity of modern systems and the need to work using methods such (p.143) "as agile, scrum and kanban [that] help… a lot in addressing the change and delivering value with a quick impact and continual basis." The authors go on to note that there is still work to be done to integrate system modelling and agile tools.

Similar issues have been traversed in work conducted for multiple agencies working under the US Department of Defense (DoD) and reported by a team led by Boehm (2018). They too were seeking to develop novel ways to reconcile model-based approaches to specifying system behaviours with increasingly agile development processes and practices in the DoD. The utility of continuous V&V when integrated with MBSE has also been considered in naval air systems (Giammarco et al., 2018) and for complex building and civil infrastructure construction projects was also considered by Chen and Jupp (2018), with the latter authors noting positive information management outcomes.

*Agile for safety- and security-critical systems*

Across two similar papers and a book Boehm and colleagues (Boehm et al., 2019a; 2019b; Rosenberg et al., 2020) describe the development of an approach they call Parallel Agile (PA) as being more suitable (Boehm et al., 2019a, p.1) "…for projects involving multiple organizations, over-100 developers, safety- and security-critical systems, and interoperability with independently-evolving systems." In particular, PA uses a 'Three-Team' approach to intentionally support either incremental or continuous development and deployment enabling rapid change *and* high assurance (Figure 1).

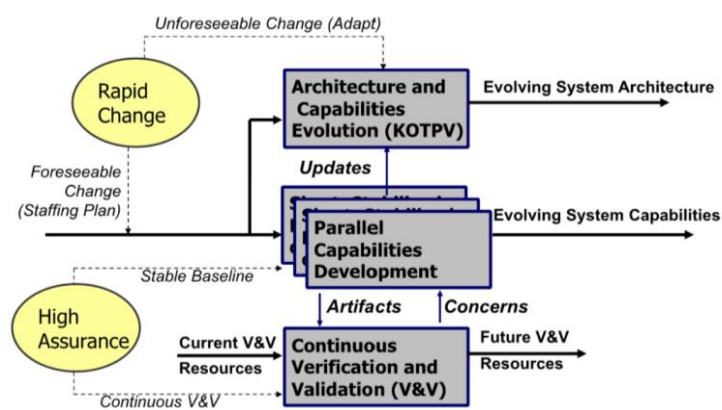

Figure 1. The Parallel Agile Three-Team Approach proposed by Boehm et al. (2019a)

Of particular note is that the work of the Continuous V&V team is just that – continuous. The team (p.3) "…does not wait for some code to be tested. It starts at the beginning of the project getting ready to support the [Incremental Commitment Spiral Model] ICSM principle and PA Critical Success Factor "Evidence and Risk-Based Decisions" by evaluating the feasibility of the initial decisions of the first two teams… As shortfalls in feasibility evidence are uncertainties and probabilities of loss, and Risk Exposure is calculated as (Probability of Loss) times (Impact of Loss), such evaluations enable the first two teams to explore less risky alternative solutions. The Continuous V&V team continues to prepare for and perform testing and evaluation of the first two teams' later artifacts." (Hagemann et al. (2020) also promote the continuous monitoring of system states following (p.216) "…a real shift left of verification and validation (V & V) activities in model-driven development processes.") Boehm et al. (2019a) also report on the key role played by the human-in-the-loop, the so-called Keeper Of The Project Vision (KOTPV), who in fact may be more than one person (p.2): "…often a combination of a domain expert and a technology expert" and may even need to be a team of experts.

An alternative approach called 'Scrum for Safety' (S4S) has been proposed more recently by Carbone et al. (2021), with an eye towards the railway domain. In building toward their own work the authors provide a useful review of relevant prior literature and address some apparent obstacles impeding the adoption of agile methods in the context of safety-critical systems. Specifically, these are the need for documentation and the need for testing to be conducted late and not by the developers themselves. (The same issues were identified in a systematic literature review reported by Heeager and Nielsen (2018), alongside requirements flexibility and process iteration.) In summary, they report growing empirical evidence that refutes both of those obstacles. They cite two particular methodologies, R-Scrum and SafeScrum, noting that (p.129) "…[t]he latter represents the result of a theoretical work in which Scrum has been brought into compliance with various standards in the critical systems world, including the IEC 61508… and the CENELEC EN 50128. Recently, this agile development methodology has been profitably used in a variety of contexts, including military…, railway… and aerospace…" (SafeScrum is fully described by Hanssen et al. (2018) and functional safety testing and V&V with agile methods are addressed by Myklebust and Stålhane (2021).)

S4S has also been designed intentionally to align with the principles and values of Scrum without compromising safety, as per the requirements described in the CENELEC EN50126, EN50128 and EN50129 standards. The approach is reflected in the following principles (Carbone et al. (2021)):

1. Cover all the alternatives before making some decision.
2. Experiment and fail frequently.
3. Deliver software continuously to the users.
4. Integrate software continuously with other actors.
5. Continuously V&V (Verify and Validate).
6. Make your work traceable.
7. Let your approach be risk-based.
8. Don't break or lose the already achieved quality.

The S4S workflow is depicted in Figure 2. While it will require further, independent evaluation the authors did assess the utility of their approach by applying it to a first case study (p.135) "…on a very highly complex research product with changeable requirements… a real-world safety-critical product, owned by one of the most important Italian company [*sic*] in the railway domain." Notably the study employed four-week sprints to specifically accommodate software quality assurance. Through this study the authors stated that they verified that risks, hazards and side-effects were adequately covered; that wasted documentation rework was avoided; that continuous testing supported by automated testing tools enabled more timely discovery and fixing of errors; and that in regard to requirements change (p.137) "…even inside a strongly regulated environment, the agile core mindset of S4S demonstrated to remain an effective tool for uncertain problem solution. Thanks to frequent reviews and releases, our customer and other involved teams became an integrative part of the development workflow, contributing to stabilizing software requirements. Also, by not planning all future sprints, we always reserved the possibility to accept some new features and technological changes, absorbing market trends. However, for each change, we had to pay a cost to update our formal specifications, which had to reflect the research state of work, in order to enable successive quality assurance activities." They credited their continuous V&V and traceability activities as in fact incrementally increasing team trust in the quality of the product.

In addressing the need for certain aspects of a system's performance to be certified Dupont et al. (2021) illustrate the potential of a proposed lightweight and flexible incremental certification process that can be integrated with a DevSecOps way of working to ensure frequently updated products and services can be certified as cybersecure (see Figure 3).

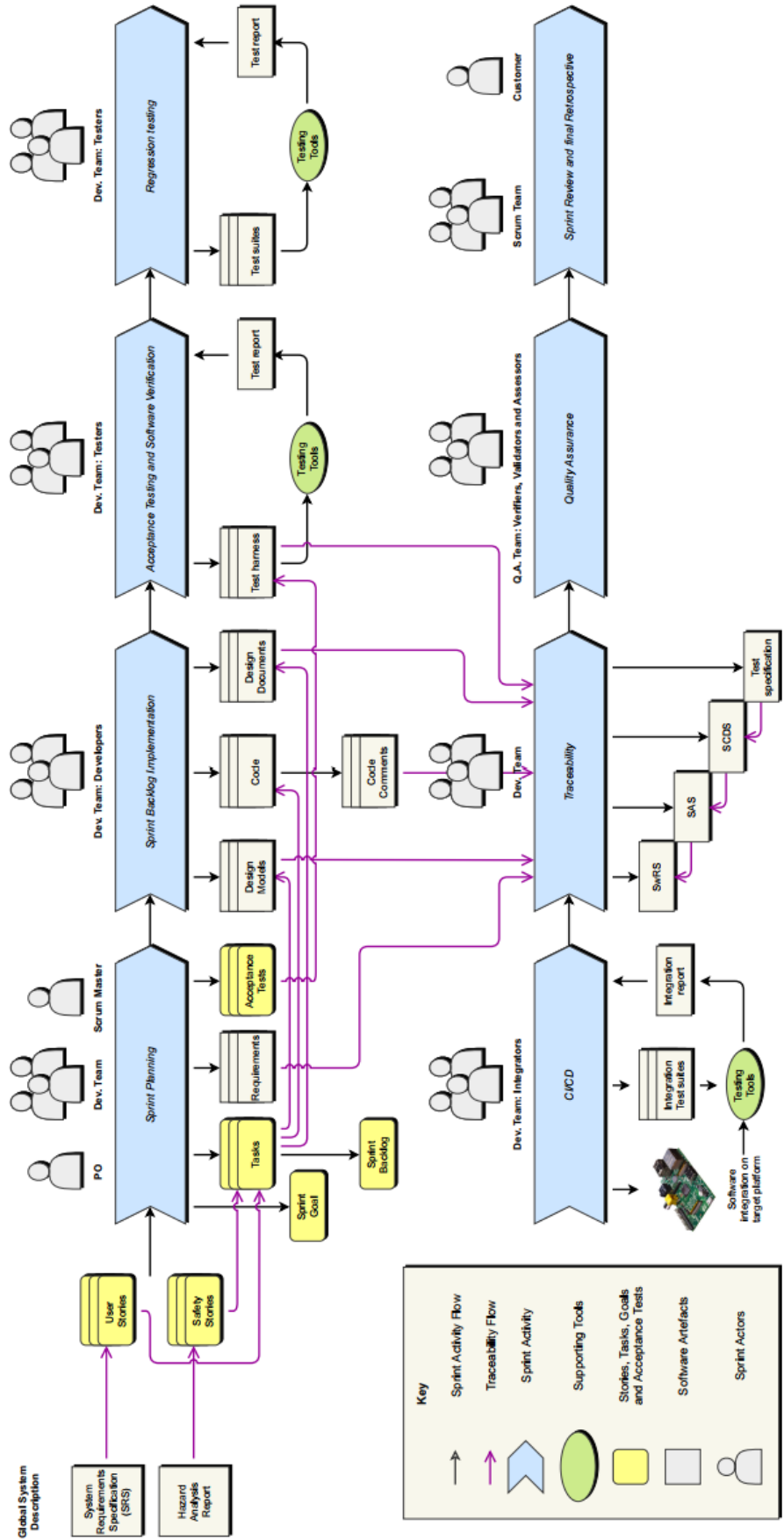

Figure 2. The Scrum for Safety Workflow proposed by Carbone et al. (2021)

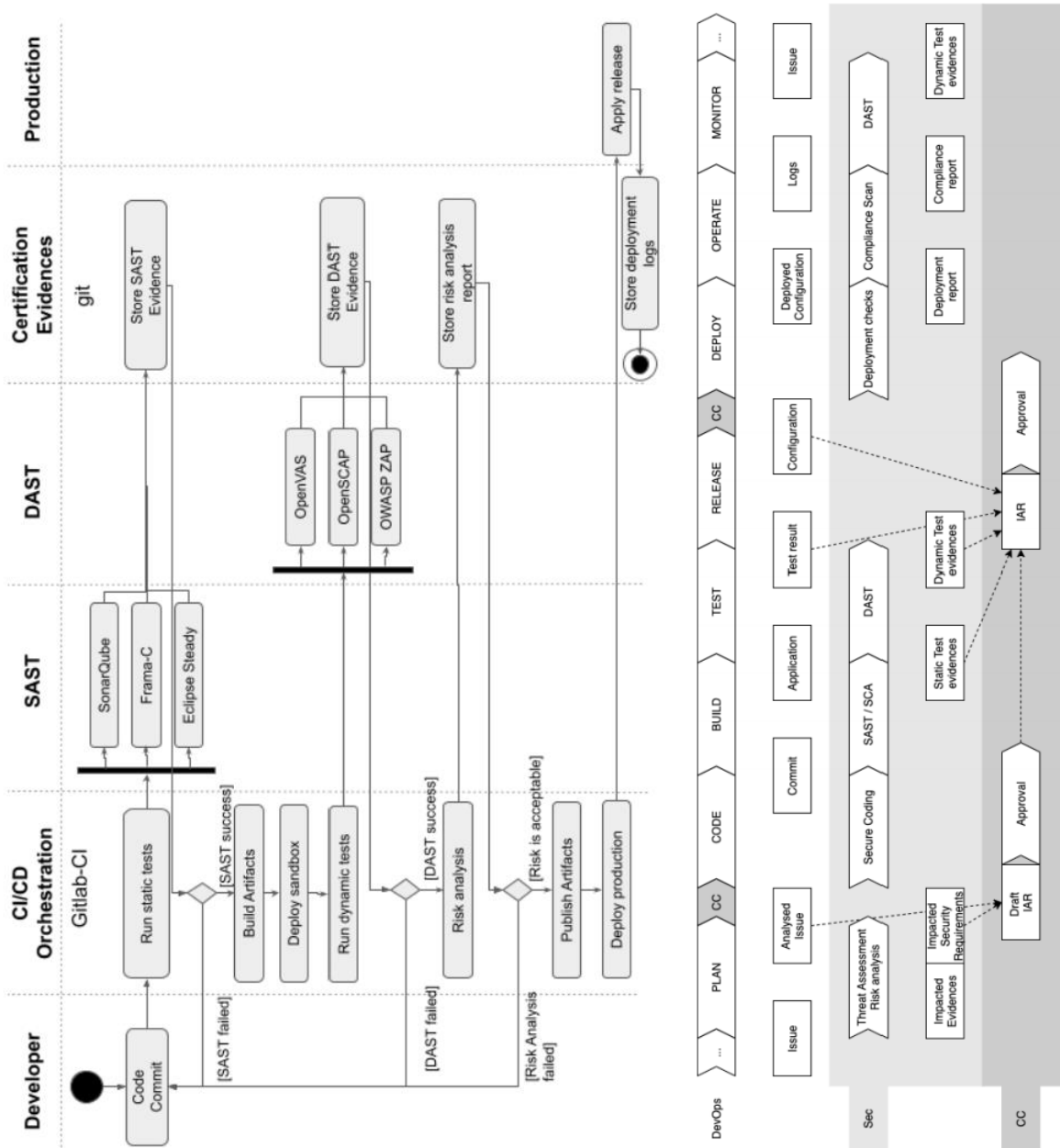

Figure 3. (a) DevSecOps process and tools (at left)
(b) overlaid with incremental certification (at right) (Dupont et al., 2021)

Even NASA has had to take the bold step of embracing agile approaches to development and delivery while continuing to use Independent Verification and Validation (IV&V) to assure the safety-critical nature of their Orion spacecraft software. Smith et al. (2019) cite the following as rationale (p.1): "Assuring the safety and performance of the embedded flight software is quickly growing beyond the reach of traditional methods and resource levels. The methods used to build these software-dominant systems evolve in an on-going attempt to keep pace with the scope of our ambitions. Agile software development is now commonplace. The long timelines and large batches of work associated with traditional methods are being replaced by rapid delivery of small increments – as system capabilities are realized in waves… Widening our aperture to encompass a dramatically larger mission scope, while adjusting our cadence to synchronize with the rapid pace of agile software development, a new approach to IV&V is emerging. with an approach matched to the evolving development methods".

The amount of work required of NASA's Orion IV&V team was outstripping their ability to deliver across the range of safety-critical domains. In addition (p.2) "…Orion IV&V leadership had a sense that the team was wasting some of its effort on assuring relatively unimportant things due to the coarseness of its scoping approach which treated an entire domain of the software as either in-scope or out-of-scope." This led to the creation of a new approach that enabled the team to focus their limited IV&V attention on the areas of highest risk in terms of the safety- and mission-critical software capabilities embedded in the Orion spacecraft. This approach, referred to as Follow-the-Risk Capability Based Assurance, is depicted in Figure 4.

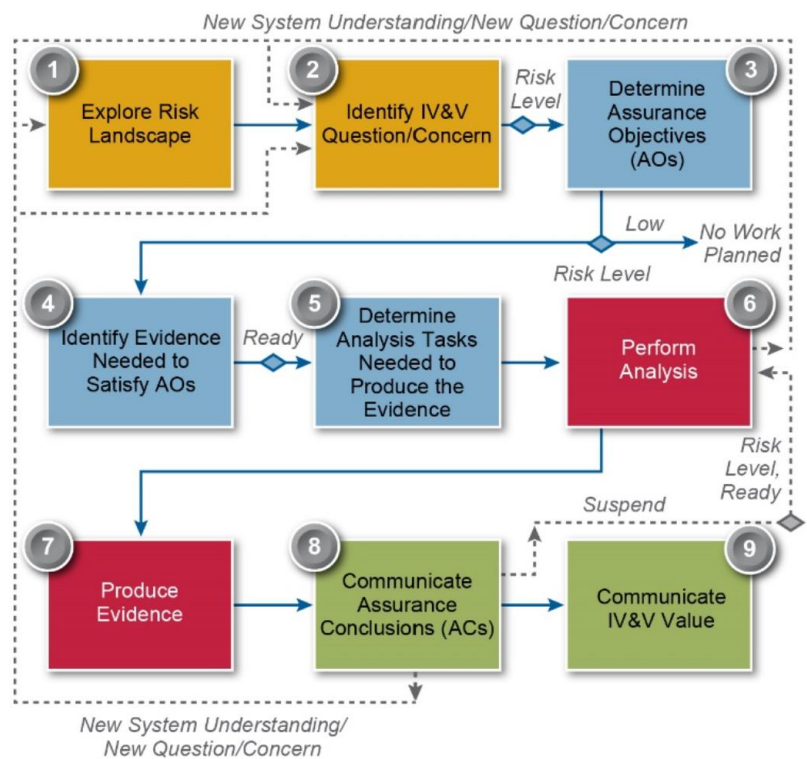

Figure 4. The Follow-the-Risk approach adopted by the NASA Orion IV&V team (Smith et al., 2019)

Smith et al. (2019) noted that this was a departure from the standards-based approach they had been using to date (p.3): "This was a very different approach than IV&V was used to in the past. This capability based assurance approach would take IV&V away from a more traditional IV&V approach derived from the IEEE 1012 Verification and Validation standard [1], to looking at artifacts and doing analysis with a specific mission capability, and the system and software capabilities that enable it, in mind." Far from jeopardising their ability to deliver, the new approach has produced benefits and has been met with support (p.8): "All levels of stakeholders, from NASA IV&V leadership up through NASA's Office of Safety and Mission Assurance and the Orion Program are extremely impressed by the quality of the work that Orion IV&V is performing." The value-add has come in terms of: better identifying evidence-based risk-driven critical issues while not inundating the development team with low-impact and trivial issues; the provision of positive assurance that confirms the flight software's "goodness"; and doing both of those tasks (pp.8-9): "…at a much faster cadence. Under the old approach it was noted that IV&V was delivering products months out of phase with the developer. Since the changes have been made we have improved our delivery cadence from months to weeks, becoming more in sync with the developer which is crucial to make IV&V worth the investment when dealing with a project using agile development methods."

*Hybrid agile and IV&V*

As identified in the findings of the international HELENA study, a common compromise sought by organisations struggling to adapt to an agile way of working, in what are perceived to be incompatible contexts, is to use a hybrid of methods. Dabney and Arthur (2019) note that this might be useful in particular for organisations undertaking large multiyear developments of mission-critical systems. This presents its own challenges for IV&V, however. Drawing on the relevant literature and guided by their experience the authors assigned IV&V techniques into three groups: thirteen techniques that could be applied as defined or 'as is' within a hybrid-agile method; ten that could be "readily tailored" (p.353) to be compatible; and a remaining group of seven techniques focused on completeness and emergent behaviour that the authors say are incompatible with a hybrid agile approach. Helpfully, they proceed to pare back those techniques to identify their underlying objectives and then propose alternative means to achieve those same objectives in a manner compatible with an agile way of working.

*Conclusions*

Software-based technologies continue to advance at pace, as do the ways in which software products and services are designed, built and deployed. As a universal technology, software is used in every domain and at every scale, from simple entertainment apps through to the critical infrastructure on which economies and societies depend. As such we need software to work, to behave and perform as we need it to. We want it to be predictable. But we also need it to change, sometimes quickly. And always be secure, yet be easy to learn, and be usable for all… Assuring that all of these (in part conflicting) aspects of quality are constantly delivered is therefore a complex undertaking, requiring multiple strategies, methods, techniques and tools. Together they can give us a known level of confidence of a system's performance, and when this is supported by human-in-the-loop foundations and sound risk management and mitigation, we have the basis for the state of the art in terms of safe, sound quality assurance in the context of contemporary practice.

*Author bio:* Stephen G. MacDonell is Professor of Software Engineering at the Auckland University of Technology and Professor in Information Science at the University of Otago, both in New Zealand. Stephen was awarded BCom(Hons) and MCom degrees from the University of Otago and a PhD from the University of Cambridge. His research has been published in *IEEE Transactions on Software Engineering*, *ACM Transactions on Software Engineering and Methodology*, *ACM Computing Surveys*, *Empirical Software Engineering*, *Information & Management*, the *Journal of Systems and Software*, *Information and Software Technology*, and the *Project Management Journal*, and his research findings have been presented at more than 100 international conferences. He is a Fellow of IT Professionals NZ, Senior Member of the IEEE and the IEEE Computer Society, Member of the ACM, and he serves on the Editorial Board of *Information and Software Technology*. He is also Deputy Director of New Zealand's National Science Challenge *Science for Technological Innovation*, Technical Advisor to the Office of the Federation of Māori Authorities Pou Whakatāmore Hangarau - Chief Advisor Innovation & Research, and Deputy Chair of Software Innovation New Zealand (SI^NZ).

*References*

1. Baduel, R., Chami, M., Bruel, J-.M., and Ober, I. (2018) SysML Models Verification and Validation in an Industrial Context: Challenges and Experimentation, in A. Pierantonio and S. Trujillo (Eds.), *Proceedings of the European Conference on Modelling Foundations and Applications (ECMFA)*, LNCS 10890, Springer, pp.132–146.
2. Beck, K., Beedle, M., van Bennekum, A., Cockburn, A., Cunningham, W., Fowler, M., Grenning, J., Highsmith, J., Hunt, A., Jeffries, R., Kern, J., Marick, B., Martin, R.C., Mellor, S., Schwaber, K., Sutherland, J., and Thomas, D. (2001) *Manifesto for Agile Software Development* https://www.agilemanifesto.org/
3. Boehm, B. (2018) *System Qualities (SQs)* Ontology, Tradespace and Affordability (SQOTA), Phase 6: 2017-2018, *Technical Report SERC-2018-TR-108*, Systems Engineering Research Center, 48p.
4. Boehm, B., Rosenberg, D., and Siegel, N. (2019a) Critical Success Factors for Scaling Agile Development, submitted to the *International Conference on Software and System Processes*, pp.1-8.
5. Boehm, B., Rosenberg, D., and Siegel, N. (2019b) Critical Quality Factors for Rapid, Scalable Agile Development, in *Proceedings of the IEEE 19th International Conference on Software Quality, Reliability and Security Companion (QRS-C)*, pp.514-515.
6. Borky, J.M., and Bradley, T.H. (2019) *Effective Model-Based Systems Engineering*, Springer.
7. Brooks, F.P. (1986/1987) No Silver Bullet – Essence and Accident in Software Engineering, *IEEE Computer* 20(4), pp.10-19.
8. Bruel, J-M., Mazzara, M., and Meyer, B. (2019) *Software Engineering Aspects of Continuous Development and New Paradigms of Software Production and Deployment*, First International Workshop, DEVOPS 2018, Springer.
9. Carbone, R., Barone, S., Barbareschi, M., and Casola, V. (2021) Scrum for Safety: Agile Development in Safety-Critical Software Systems, in A. C. R. Paiva et al. (Eds.), *Proceedings of the International Conference on the Quality of Information and Communications Technology (QUATIC),* CCIS 1439, pp.127–140.
10. Carrozza, G., Pietrantuono, R. and Russo S. (2018) A software quality framework for large-scale mission-critical systems engineering, *Information and Software Technology* 102, pp.100-116.
11. Chen, Y., and Jupp, J. (2018) Model-Based Systems Engineering and Through-Life Information Management in Complex Construction, in P. Chiabert et al. (Eds.), *Proceedings of the Conference on Product Lifecycle Management (PLM),* IFIP AICT 540, Springer, pp.80–92.
12. Cheriyan, A., Gondkar, R.R., and Babu, S.S. (2020) Quality Assurance Practices and Techniques Used by QA Professional in Continuous Delivery, in M. Tuba et al. (Eds.), *Information and Communication Technology for Sustainable Development,* Advances in Intelligent Systems and Computing 933, Springer, pp.83-92.
13. Corral, L., and Fronza, I. (2018) Design Thinking and Agile Practices for Software Engineering: An Opportunity for Innovation, in *Proceedings of the 19th Annual SIG Conference on Information Technology Education (SIGITE),* ACM Press, pp. 26-31.
14. Dabney, J.B., and Arthur, J.D. (2019) Applying Standard Independent Verification and Validation Techniques within an Agile Framework: Identifying and Reconciling Incompatibilities, *Systems Engineering* 22, pp.348-360.
15. Digital.ai (2021) *15th State of Agile Report*, Digital.ai, 23p.
16. Drechsler, A., and Breth, S. (2019) How to Go Global: A Transformative Process Model for the Transition Towards Globally Distributed Software Development Projects, *International Journal of Project Management* 37, pp.941–955.

17. Dupont, S., Ginis, G., Malacario, M., Porretti, C., Maunero, N., Ponsard, C., and Massonet, P., (2021) Incremental Common Criteria Certification Processes using DevSecOps Practices, in *Proceedings of the IEEE European Symposium on Security and Privacy Workshops (EuroS&PW),* IEEE CS Press, pp.12-23.
18. Fitzgerald, B., and Stol, K-J., (2017) Continuous Software Engineering: A Roadmap and Agenda, *Journal of Systems and Software* 123, pp.176–189.
19. Gerostathopoulos, I., et al. (2019) Continuous Data-driven Software Engineering – Towards a Research Agenda, *ACM SIGSOFT Software Engineering Notes* 44(3), pp.60–64.
20. Giammarco, K., et al. (2018) Verification and Validation (V&V) of System Behavior Specifications, *Final Technical Report SERC-2018-TR-116*, Systems Engineering Research Center, 141p.
21. Hagemann, N., Pröll, R., and Bauer B. (2020) Towards Abstract Test Execution in Early Stages of Model-driven Software Development, in *Proceedings of the 8th International Conference on Model-Driven Engineering and Software Development (MODELSWARD)*, pp.216-226.
22. Hanssen, G.K., Stålhane, T., and Myklebust, T. (2018) *SafeScrum® – Agile Development of Safety-Critical Software*, Springer.
23. Heeager, L.T., and Nielsen P.A. (2018) A Conceptual Model of Agile Software Development in a Safety-Critical Context: A Systematic Literature Review, *Information and Software Technology* 103, pp.22-39.
24. IEEE (2017) *IEEE Standard for System, Software, and Hardware Verification and Validation,* IEEE Standards Association of the IEEE Computer Society, IEEE Std 1012™-2016 (Revision of IEEE Std 1012-2012/Incorporates IEEE Std 1012-2016/Cor1-2017).
25. IEEE (2021) IEEE Standard for DevOps: Building Reliable and Secure Systems Including Application Build, Package, and Deployment, IEEE Standards Association of the IEEE Computer Society, IEEE Std 2675™-2021.
26. Islam, G., and Storer, T. (2020) A Case Study of Agile Software Development for Safety-Critical Systems Projects, *Reliability Engineering and System Safety* 200, Article 106954, 18p.
27. Kersten, M. (2019) What Flows through a Software Value Stream? *IEEE Software*, July/August, pp.8-11.
28. Klünder, J., Hebig, R., Tell, P., Kuhrmann, M., Nakatumba-Nabende, J., Heldal, R., Krusche, S., Fazal-Baqaie, M., Felderer, M., Genero Bocco, M.F., Küpper, S., Licorish, S.A., López, G., McCaffery, F., Top, O.O., Prause, C., Prikladnicki, R., Tüzün, E., Pfahl, D., Schneider, K., & MacDonell, S.G. (2019) Catching up with method and process practice: An industry-informed baseline for researchers, in *Proceedings of the International Conference on Software Engineering - Software Engineering in Practice (ICSE-SEIP2019)*. Montréal, Canada, IEEE Computer Society Press, pp.255-264.
29. Kuhrmann, M., Nakatumba-Nabende, J., Pfeiffer, R.-H., Tell, P., Klünder, J., Conte, T., MacDonell, S.G., & Hebig, R. (2019) Walking through the method zoo: Does higher education really meet software industry demands?, in *Proceedings of the International Conference on Software Engineering - Software Engineering Education and Training (ICSE-SEET2019)*. Montréal, Canada, IEEE Computer Society Press, pp.1-11.
30. Kuhrmann, M., Tell, P., Hebig, R., Klünder, J., Münch, J., Linssen, O., Pfahl, D., Felderer, M., Prause, C., MacDonell, S.G., Nakatumba-Nabende, J., Raffo, D., Beecham, S., Tuzun, E., Lopez, G., Paez, N., Fontdevila, D., Licorish, S., Küpper, S., Ruhe, G., Knauss, E., Özcan-Top, Ö., Clarke, P., McCaffery, F., Genero, M., Vizcaino, A., Piattini, M., Kalinowski, M., Conte, T., Prikladnicki, R., Krusche, S., Coskuncay, A., Scott, E., Calefato, F., Pimonova, S., Pfeiffer, R.-H., Pagh Schultz, U., Heldal, R., Fazal-Baqaie, M., Anslow, C., Nayebi, M., Schneider, K., Sauer, S.,

Winkler, D., Biffl, S., Bastarrica, C., & Richardson, I. (Accepted, In Press) What makes agile software development agile?, *IEEE Transactions on Software Engineering*, pp.TBC.
31. Konersmann, M., Fitzgerald, B., Goedicke, M., Olsson, H.H., Bosch, J., and Krusche, S. (2020) Rapid Continuous Software Engineering – State of the Practice and Open Research Questions, *ACM SIGSOFT Software Engineering Notes* 46(1), pp.25–27.
32. KPMG (2019) *Agile Transformation*, KPMG Advisory N.V., 42p.
33. Liang, H-.H., Ye, W-.P., Liu, W., and Tang J-.Z. (2021) Discussion on the Software V&V Technology in Nuclear Power Plants, Y. Xu et al. (Eds.), *Nuclear Power Plants: Innovative Technologies for Instrumentation and Control Systems*, LNEE 779, pp.380–385.
34. Myklebust, T., and Stålhane, T. (2021) *Functional Safety and Proof of Compliance*, Springer.
35. Pröll, R. (2021) *Towards a Model-Centric Software Testing Life Cycle for Early and Consistent Testing Activities*, Dissertation for the degree of Doctor of Natural Sciences, University of Augsburg, Augsburg, Germany.
36. Puntigam, W., Zehetner, J., Lappano, E., and Krems, D. (2020) Integrated and Open Development Platform for the Automotive Industry, in H. Hick et al. (Eds.), *Systems Engineering for Automotive Powertrain Development*, Springer, pp.471-497.
37. Reiher, D., and Hahn, A. (2021) Review on the Current State of Scenario- and Simulation-Based V&V in Application for Maritime Traffic Systems, in *Proceedings of OCEANS 2021*, IEEE CS Press, pp.1-9.
38. Risks Digest (n.d.) *Forum on Risks to the Public in Computers and Related Systems,* Peter G. Neumann – Moderator https://catless.ncl.ac.uk/risks/
39. Rosenberg, D., Boehm, B., Stephens, M., Suscheck, C., Dhalipathi, S.R., and Wang, B. (2020) *Parallel Agile – Faster Delivery, Fewer Defects, Lower Cost*, Springer.
40. Royce, W.W. (1987) Managing the development of large software systems, in *Proceedings of the 9th international Conference on Software Engineering (ICSE),* pp.328–338.
41. Sauer, C., and Reich, B.H. (2009) Rethinking IT Project Management: Evidence of a New Mindset and its Implications, *International Journal of Project Management* 27, pp.182–193.
42. Shameem, M., Kumar, R.R., Kumar, C., Chandra, B., and Khan, A.A. (2018) Prioritizing Challenges of Agile Process in Distributed Software Development Environment Using Analytic Hierarchy Process, *Journal of Software: Evolution and Process* 30:e1979, 19p.
43. Smith, J., Bradbury, J., Hayes, W., and Deadrick, W. (2019) Agile approach to assuring the safety-critical embedded software for NASA's Orion spacecraft, *in Proceedings of the International Conference for Aerospace Experts, Academics, Military Personnel, and Industry Leaders*, 10p.
44. Svejvig, P., and Andersen, P. (2015) Rethinking Project Management: A Structured Literature Review with a Critical Look at the Brave New World, *International Journal of Project Management* 33, pp.278–290.
45. Tahera, K., Wynn, D.C., Earl, C., and Eckert, C.M. (2019) Testing in the Incremental Design and Development of Complex Products, *Research in Engineering Design* 30, pp.291–316.
46. Tell, P., Klünder, J., Küpper, S., Raffo, D., MacDonell, S.G., Münch, J., Pfahl, D., Linssen, O., & Kuhrmann, M. (2019) What are hybrid development methods made of? An evidence-based characterization, in *Proceedings of the International Conference on Software and System Processes (ICSSP2019)*. Montréal, Canada, IEEE Computer Society Press, pp.105-114.
47. Tell, P., Klünder, J., Küpper, S., Raffo, D., MacDonell, S.G., Münch, J., Pfahl, D., Linssen, O., & Kuhrmann, M. (2021) Towards the statistical construction of hybrid development methods, *Journal of Software: Evolution and Process* 33(e2315), pp.1-20.

48. Ugarte Querejeta, M., Etxeberria, L., and Sagardui G. (2020) Towards a DevOps Approach in Cyber Physical Production Systems Using Digital Twins, in A. Casimiro et al. (Eds.), *SAFECOMP 2020 Workshops*, LNCS 12235, pp.205–216.
49. Wolpers, S. (2021) *The Agile Metrics Survey 2021*
https://www.scrum.org/resources/blog/agile-metrics-survey-2021
50. Yoder, J.W., Wirfs-Brock, R., and Washizaki, H. (2015) QA to AQ Part Four: Shifting from Quality Assurance to Agile Quality "Prioritizing Qualities and Making them Visible", in *HILLSIDE Proceedings of Workshops at the 22<sup>nd</sup> Conference on Pattern Languages of Programs (PLoP),* 14p.

*Appendix 1*

Pröll (2021) provides the following overview of standards relevant to software testing and quality assurance, illustrating the multiple parties and perspectives involved (p.44):

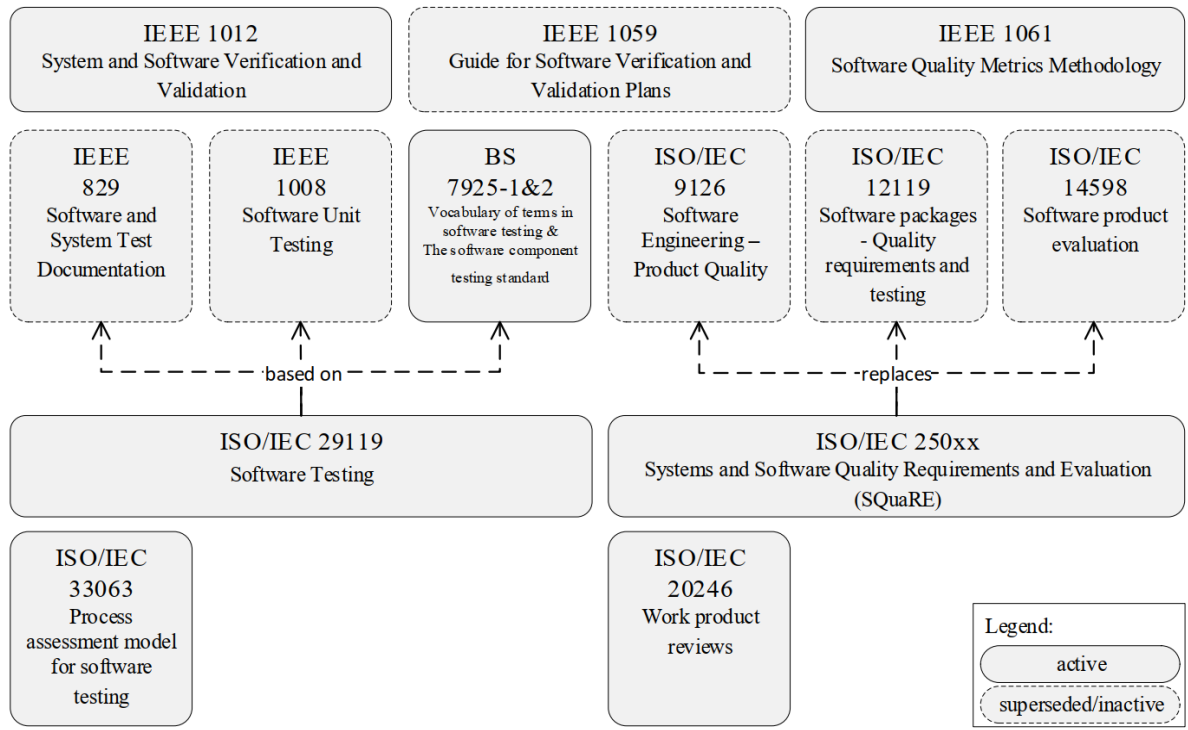